# Hybrid membrane resonators for multiple frequency asymmetric absorption and reflection in large waveguide

Caixing Fu, Xiaonan Zhang, Min Yang, Songwen Xiao, and Z. Yang*
Department of Physics, Hong Kong University of Science and Technology, Clearwater Bay, Kowloon, Hong Kong SAR, China

**Abstract**

We report that Hybrid membrane resonators (HMRs) made of a decorated membrane resonator backed by a shallow cavity can function as Helmholtz resonators (HRs) when mounted on the sidewall of a clear waveguide for air ventilation. When two single-frequency HMRs are used in the same scheme as two frequency-detuned HRs, asymmetric total absorption/reflection is demonstrated at 286.7 Hz with absorption coefficient over 97 % in a waveguide 9 cm × 9 cm in cross section. When two multiple-frequency HMRs are used, absorption in the range of near 60 % to above 80 % is observed at 403 Hz, 450 Hz, 688 Hz, 863 Hz and 945 Hz. Theoretical predictions agree well with the experimental data. The HMRs may replace HRs in duct noise reduction applications in that at a single operation frequency they have stronger strength to cover a much larger cross section area than that of HRs with similar cavity volume, and they can be designed to provide multiple frequency absorption band.

Total absorption in linear dissipative systems within the subwavelength scale had been a challenge until recently [1 - 8]. Several schemes have been theoretically proposed via coherent perfect absorber (CPA), where two coherent waves with specific amplitude and phase were incident in opposite direction to an absorption core [1, 2]. For one beam incidence scenario, maximum absorption by a single dipole or monopole unit was shown to be at most 50 % [3]. Dark acoustic metamaterials backed by a hard wall could achieve total absorption via curvature energy at resonance to maximize energy dissipation while achieving impedance match to eliminate reflection [4], while meta-surface made of hybrid resonators [5] could provide an effective perfect absorption area several times larger than the physical area of the active devices. Perfect absorption by a meta-surface consisting of a perforated plate and a coiled coplanar air chamber has also been demonstrated [6]. These types of devices all contained a hard wall to eliminate transmission. The hard wall was replaced in a monopole-dipole co-resonance scheme to eliminate transmission, while the individual resonances ensure impedance matching and elimination of reflection [7]. No airflow was allowed in the direction of wave propagation, except for the second device reported in Ref. 7, but the dipole sub-component blocked part of the waveguide cross section so it still impeded airflow. Most recently, two Helmholtz resonators (HRs) mounted on the sidewall of a waveguide were reported to have nearly perfect absorption at low frequency [8]. For devices with physical sizes several times the relevant wavelength, perfect absorption could be achieved by slowing down the waves [9, 10] and in broadband [11]. However, up to now, no broadband sub-wavelength absorbers mounted on the sidewall of clear waveguide have been experimentally realized.

In this letter, we report ventilated perfect absorbers comprising two sub-wavelength hybrid membrane resonators (HMRs) which form part of the sidewall in line with the rest of the ventilation waveguide. For two single-resonant-frequency HMRs with one being slightly detuned from the other, total absorption is achieved when waves with a frequency somewhere in-between the two HMR resonant frequencies are incident from the high resonant frequency HMR side, while total reflection is achieved if the waves are incident from the other side. For two multiple-resonant-frequency HMRs with slightly detuned resonant frequencies from one another, strong asymmetric absorption and reflection are experimentally realized at five frequency bands below 1000 Hz.

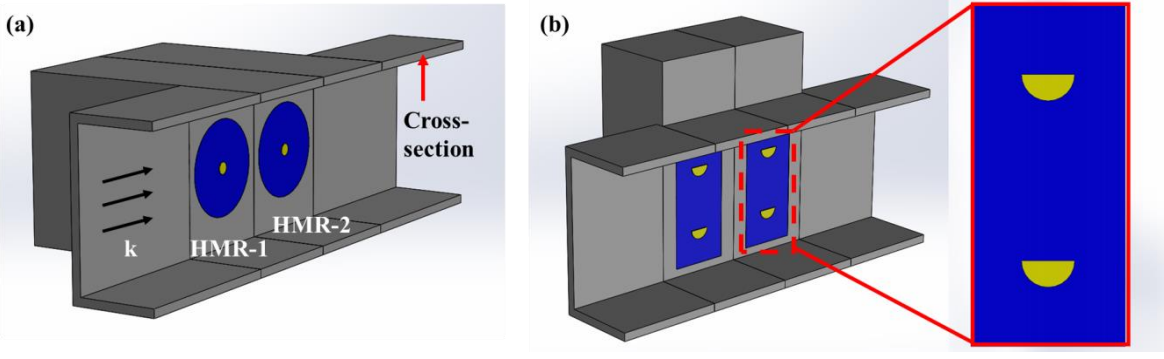

**Figure 1:** (a) A schematic view of the composite absorber that consists of two single-platelet HMRs denoted by HMR-1 and HMR-2 that form part of the sidewall of a square waveguide. The arrows denote the incident sound waves. (b) A schematic view of two two-platelet HMRs. The enlarged section of a DMR was drawn to scale as the real sample.

Consider first a decorated membrane resonator (DMR) backed by a sealed cavity, forming a HMR [5]. The surface response function of the DMR can be expressed in terms of its eigenmodes [12],

$$G_m = \sum_n \frac{|<W_n>|^2}{\rho_n (\omega_n^2 - \omega^2 + i\omega\beta_n)} \quad (1)$$

where $\rho_n = \frac{1}{A_m}\int \rho |W_n|^2 d\Omega$ is the displacement-weighted mass density, $\rho$ is the local mass density, $A_m$ is the area of the decorated membrane, $\omega_n$ is the angular frequency of the *n-th* resonant mode of the DMR, and $\omega$ is the angular frequency of the excitation. The dissipation coefficients $\beta_n$ are fitting parameters. Usually only the eigenmodes close to the frequency of interest are included in Eq. (1). The acoustic impedance of the HMR cavity is given as $Z_c = -i\gamma p_0/(V\omega)$ [5], where $\gamma$ is the adiabatic index, $p_0$ is the atmospheric pressure, and $V$ is the volume of the cavity. The total acoustic impedance of the HMR is $Z_r = Z_m + Z_c$, where $Z_m^{-1} = i\omega G_m A_m$. When the HMR is mounted on the sidewall of a waveguide, as shown in Fig. 1(a), the acoustic impedance across the waveguide (waveguide impedance) is $Z^{-1} = Z_r^{-1} + Z_0^{-1}$, where $Z_0 = \rho_0 c_0/A$ is the acoustic impedance of air in the waveguide, $\rho_0$ is the density of air, $c_0$ is the speed of sound and $A$ is the cross section area of the waveguide. Similar to a side-mounted HR [8], when the frequency reaches its resonance, the HMR will generate a monopole resonance that pushes or sucks the air in the axial direction along the waveguide through the normal movement of the membrane. The waveguide impedance of an HMR behaves like a soft boundary with near zero impedance, leading to minimum transmission and maximum reflection.

When two HMRs are mounted on the sidewall of the waveguide as shown in Fig. 1(a), and based on the impedance transfer method [13], the waveguide impedance in front of the second HMR (HMR-2) can be transferred to the surface in front of the first HMR (HMR-1). The combined waveguide impedance (CWI) at HMR-1 is then

$$Z_{CW1}^{-1} = \frac{1}{Z_0} \frac{Z_0 + iZ_{r2}\tan(\omega L/c_0)}{Z_{r2} + iZ_0\tan(\omega L/c_0)} + Z_{r1}^{-1} \tag{2}$$

where $Z_{r1}$ and $Z_{r2}$ denote the acoustic impedance of HMR-1 and HMR-2 given by its response function of DMR and cavity, and $L$ is the distance between the two HMRs measured from their centers. The CWI at HMR-2 ($Z_{CW2}$) can be obtained by exchanging $Z_{r1}$ and $Z_{r2}$ in Eq. (2).

Let the resonant frequencies of the two HMRs be $f_1$ and $f_2$, with $f_1$ being slightly higher than $f_2$. For waves incident from the HMR-1 side, at a frequency near $f_2$ the impedance of HMR-2 is almost zero, so $Z_{CW1}^{-1} \approx \frac{1}{iZ_0\tan(\omega L/c_0)} + Z_{r1}^{-1}$. It is seen from the expression that HMR-2 generates a perfect-reflection boundary in the waveguide, and the air column between the two HMRs serves as a cavity that is in parallel to HMR-1. Similar to the case of a HMR [5], it is possible for $Z_{CW1}$ to be equal to $Z_0$, leading to zero reflection. As the transmission is also small, this leads to total absorption of the incident waves. For waves incident from the HMR-2 side at the same frequency, $Z_{r2}$ is near zero so $Z_{CW2} \approx Z_{r2} \approx 0$, which leads to a soft boundary and total reflection. Therefore, when the waves at frequency close to $f_2$ are incident from the HMR-1 side, near total absorption will occur. Near total reflection will occur if the same waves are incident from the HMR-2 side. As DMR's usually have several resonant modes, especially for those with multiple platelets [4] shown in Fig. 1(b), one expects that asymmetric absorption and reflection could occur at multiple frequencies.

To understand clearly the mechanism of asymmetric absorption/reflection of HMRs we first investigate a pair of HMRs, each having just one resonance in the frequency range of interest, and demonstrate asymmetric absorption/reflection at a single frequency. The HMRs were mounted on the sidewall of a waveguide at 65 mm distance measured from center to center (Fig. 1(a)). The cross-section of the waveguide was 90 mm × 90 mm. Each HMR had a rectangular cavity that was 95 mm in depth and a 90 mm × 90 mm front surface with an opening of 55 mm in diameter. The opening was sealed by a rubber membrane of 0.15 mm in thickness and 55 mm in diameter. In HMR-2, a circle platelet with 7 mm in diameter and 420 mg in mass was attached to the center of the membrane. The mass of the platelet (the same diameter as that in HMR-2) in HMR-1 was 338 mg, and a 5 mm thick sponge was placed in the back of its cavity to tune the loss. The reflection and the transmission from the waveguide section in front the two HMRs were measured in the same way as in Ref. 7. The CWI at the surface on the incident side can be retrieved through the experimental reflection $R$ via $Z_{eq} = Z_0 \times (1 + R)/(1 − R)$.

The eigenmodes were found by COMSOL Multiphysics numerical simulations using the following parameters. The mass density, Poisson's ratio, Young's modulus, and pre-stress of the membrane were 980 $kg/m^3$, 0.49, $5 \times 10^5$, and $0.4 MPa$, respectively. The platelets are much more rigid than the membrane. The speed and mass density of air were 343 $m/s$ and 1.29 $kg/m^3$. Only the lowest resonant mode was considered for each HMR. The resonance found in the simulations for HMR-1 was at 291 Hz, and that for HMR-2 was at 286 Hz. The dissipation coefficients $\beta$ of the two HMRs were first estimated by the experimental transmission line width for each individual device, and then fine-tuned to best match the experimental results. The final values used in the simulations depicted in Fig. 2 were 23.5 Hz for HMR-1 and 13.8 Hz for HMR-2, respectively. Both are comparable with the experimental transmission line width. The theoretical transmission ($T$) was obtained from the 2 × 2 transfer matrix elements $T_{ij}$ via the expression [14]

$$T = \frac{2}{T_{11} + T_{12}/Z_0 + T_{21}Z_0 + T_{22}} \qquad (3)$$

The reflection for waves incident from the HMR-1 side and HMR-2 side can be obtained from $Z_{CW1}$ and $Z_{CW2}$, respectively. Then $\alpha = 1 - |T|^2 - |R|^2$, where $\alpha$ is the absorption coefficient.

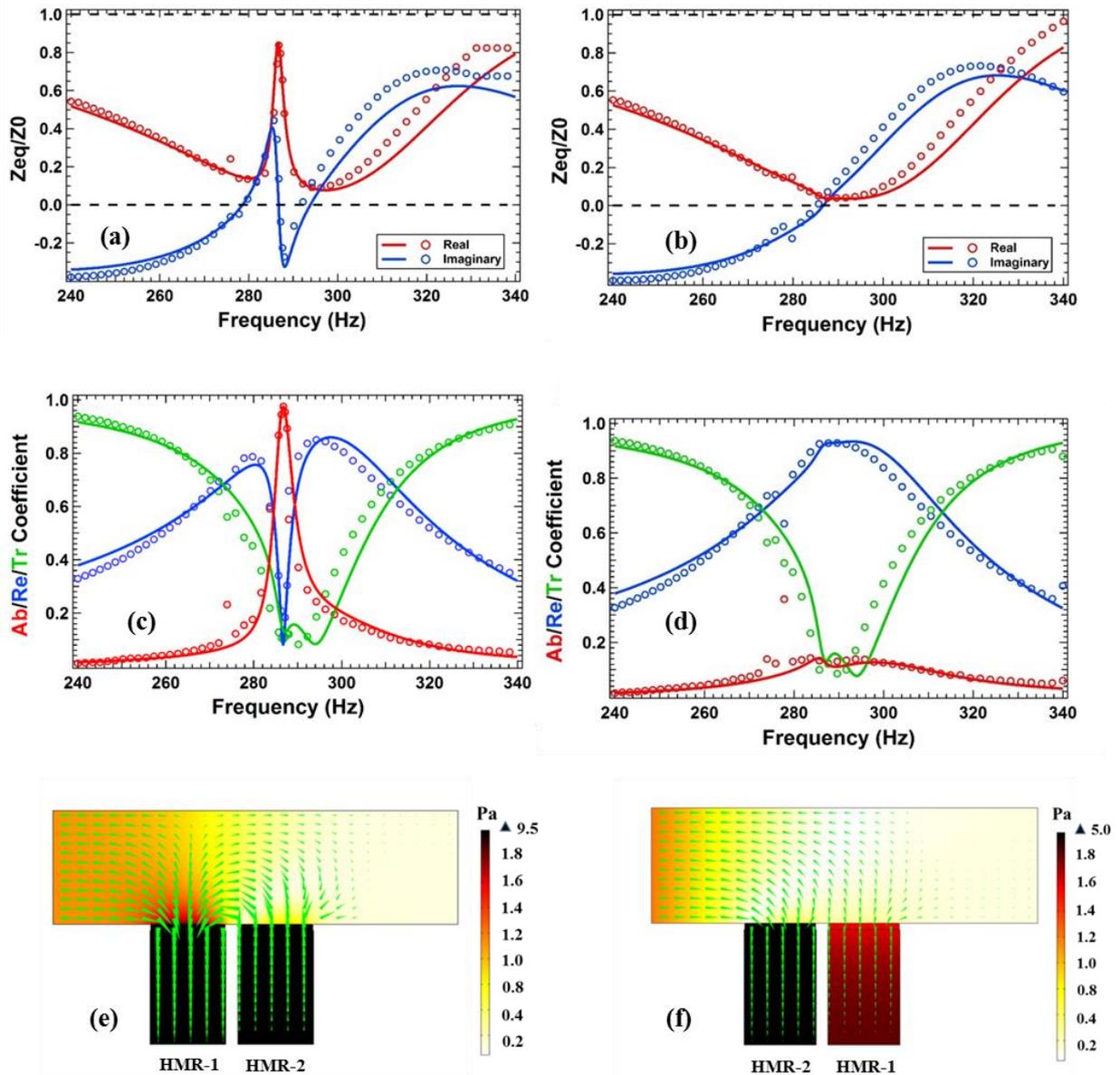

**Figure 2:** (a, b) The CWI normalized by $Z_0$ for the surface in front of (a) HMR-1 and (b) HMR-2. (c, d) The transmission (green), reflection (blue), and absorption (red) spectra when the waves are incident (c) from the HMR-1 side, and (d) from the HMR-2 side. In (a) through (d) the curves are theoretical results and the circles are experimental data. (e, f) The air velocity fields when the waves are incident (e) from the HMR-1 side, and (f) from the HMR-2 side.

The theoretical CWI $Z_{CW1}$ and $Z_{CW2}$ are shown as solid curves in Fig. 2(a) and (b), respectively. It is seen that when the waves are incident from the HMR-1 side, the real part of the CWI increases to $0.85Z_0$ while the imaginary part becomes zero at 286.7 Hz. The reflection reaches minimum due to near-match of impedance with air, as indicated by the blue curve in Fig. 2(c), which agrees well with the experimental reflection spectrum (blue circles in Fig. 2(c)). As the transmission (green circles for experimental data and green curve for theory) is small due to the monopole excitation of the HMRs, the absorption reaches the peak value of 97.5 %, as indicated by the red circles for experimental data and the red curve for theory in Fig. 2(c).

As shown in Fig. 2(b), for waves incident from the HMR-2 side, the CWI is nearly zero around 286.7 Hz, resulting in high reflection, as shown in Fig. 2(d). The transmission is identical to that for the other incident direction, as expected. In all, the theoretical results agree very well with the experimental data.

The numerical air velocity fields at the peak absorption frequency of 286.7 Hz are depicted in Fig. 2(e) and 2(f) for the two incident directions, respectively. When the wave was incident from the HMR-1 side, both HMRs were highly excited, as the air velocities inside both cavity reached over 9 times the incident wave. The directions of the air velocity near the two HMRs, however, were opposite. The incident sound waves were trapped between the HMRs and eventually dissipated by the HMRs, leading to near total absorption.

When the waves were from the HMR-2 side, only HMR-2 was excited at 286.7 Hz. The air velocity in the cavity of HMR-2 was about 5 times the incident wave, while in the cavity of HMR-1 the air velocity was only about 1.5 times the incident wave. The directions of the air velocity near the two HMRs were about the same. The wave was mostly reflected with little absorption.

For multiple frequency absorption, we made two devices HMR-3 and HMR-4. The schematics of the device structure are shown in Fig. 1(b). Both cavities were made of a 37 mm × 120 mm front plate and 40 mm in depth. An opening of 35 mm × 78 mm in dimension was made on the front plate and sealed by a rectangular membrane 0.15 mm in thickness. Two semicircle hard platelets were mounted on the membrane. The semicircle radius and the thickness were 6 mm and 0.5 mm, respectively. The mass of the top platelet in HMR-3 was 37 mg, while that of the bottom platelet was 39 mg. The corresponding ones in HMR-4 were 37 mg and 42 mg, respectively. Their exact locations on the membrane are shown in the enlarged portion of Fig. 1(b) that was drawn in scale as the real HMR-3 device. The top platelet of HMR-4 was moved up by 1 mm as compared to that of HMR-3, while the bottom platelet of HMR-4 was moved down by 1 mm as compared to that of HMR-3. The cross-section of the waveguide was 90 mm × 90 mm. A thin layer of grease was applied onto the membrane of HMR-3 to introduce extra dissipation. A single dissipation coefficient in the same value as HMR-1 and HMR-2 was used in the simulations for all the eigenmodes of HMR-3 and HMR-4, respectively. The same membrane parameters as for HMR-1 were used for HMR-3 and HMR-4. The separation between the centers of the two devices was 60 mm when both were mounted on the waveguide.

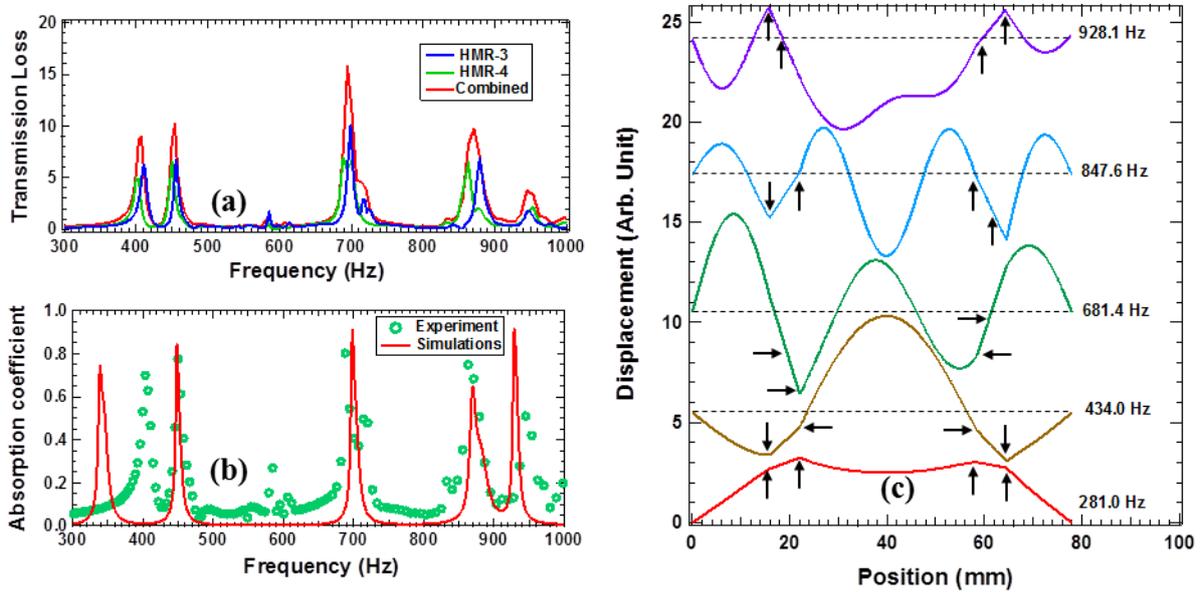

**Figure 3:** (a) Experimental TL spectra of individual HMR-3 (blue) and HMR-4 (green), and when both HMRs were used (red). (b) Absorption coefficient of the combined device made of the two HMRs. The green circles are experimental data and the red curve is the theoretical results. (c) The vibration profiles of the five eigenmodes along the long central axis of the rectangular membrane obtained by numerical simulations. The black dashed lines indicate the zero amplitude of each profile. The arrows indicate the edges of the platelets.

Figure 3(a) depicts the experimental transmission loss (TL) spectra of HMR-3 and HMR-4 when only one of them was mounted on the sidewall of the waveguide (green and blue curves), and when both were mounted in the sidewall (red curve). Five strong TL peaks can be identified for each individual HMR. Notice a small detuning of the HMR-4 TL peaks relative to those of HMR-3. The experimental absorption spectrum for the waves incident from the HMR-3 side is shown as circles in Fig. 3(b). Five strong absorption peeks can be identified at 403 Hz, 450 Hz, 688 Hz, 863 Hz and 945 Hz (Peak I – V). These absorption peaks are between the frequencies of the corresponding TL peaks of the two HMRs. Peak-II, III, and V reach above 80 %. Peak-II is near 80 %, and Peak-V is about 60 %. For the waves incident from the HMR-4 side, only strong reflection peaks were observed at the TL peak frequencies of HMR-4. The spectra are not shown here due to limit in available space.

Numerical simulations were carried out to find the eigenmodes using the device structure parameters given above. For HMR-4 five eigenmodes that produce the soft boundary were identified at 271.8, 431.8, 678.9, 841.4, and 924.2 Hz. Those for HMR-3 were at 281.0, 434.0, 681.4, 847.6, and 928.1 Hz. The vibration profiles of the five eigenmodes along the long central axis of the rectangular membrane are shown in Fig. 3(c). They all have large average displacement to generate a strong monopole resonance in the waveguide impedance. These five eigenmodes were then used in the calculations for the waveguide impedance following the same way as for HMR-1 and HMR-2. The best fit shown in Fig. 3(b) agrees well with the experimental data except for Peak-I, where the resonant frequency is off by about 120 Hz. Through theoretical investigations, we found that it is extremely difficult to optimize the delicate balance between impedance matching and dissipation cancelation for each individual absorption peak, even though we are able to match four out of the five resonant peak frequencies. What we have presented here is a preliminary success in the attempt to predict the properties of the devices, which omitted some of the structural details, such as the contact areas between the membranes and the platelets, the mass distribution of each platelet, and the stress distribution in the membrane.

In summary, we have demonstrated that an HMR functions like an HR when mounted on the sidewall of a clear waveguide, in that it also creates a soft boundary in the waveguide that causes strong reflection and transmission loss. When two slightly detuned HMRs are mounted in series on the sidewall along the waveguide without any significant impediment of airflow, asymmetric total absorption and reflection can be realized. At a single working frequency the HMRs with comparable cavity volume as the corresponding HRs have significantly stronger strength than HRs, as the waveguide cross section area in this work is over three times of that for the HRs [8]. The HMRs can also provide multiple frequency asymmetric absorption/reflection, which the HRs cannot.

Acknowledgement – We sincerely thank P. Sheng for invaluable discussions. This work was supported by AoE/P-02/12 from the Research Grant Council of the Hong Kong SAR government.